 \documentclass[12pt]{article}

\usepackage[bbgreekl]{mathbbol} 

\usepackage{amsmath,amsfonts,amssymb,amsthm}

\usepackage{slashed} 

\makeatletter
\def\@prodint#1{\vcenter{\hbox{#1$\bm{\pi}$}}}
\def\prodint{\mathop{
  \mathchoice{\@prodint\huge}{\@prodint\Large}
	{\@prodint\large}{\@prodint\normalsize}}}
\makeatother 

\DeclareSymbolFontAlphabet{\mathbb}{AMSb}
\DeclareSymbolFontAlphabet{\mathbbl}{bbold} 

\newcounter{thMM}
\newcounter{leMM}
\newcounter{deFF}
\newcounter{exMP}
\title{\large\bf  
 On the spectrum of DW Hamiltonian \\
of quantum SU(2) gauge field
\\
}
\author{ 
Igor V. Kanatchikov 
\\ 
\small\it School of Physics and Astronomy \\ 
\small\it University of St Andrews, \\ 
\small\it St Andrews KY16 9SS,  UK 
\smallskip \\ 
\small\it Quantum Information Center in Gda\'nsk (KCIK),\\ 
\small\it 81-831 Sopot, Poland  
}

\date{}

\catcode `\@=11
\@addtoreset{equation}{section}

\def\section{\@startsection {section}{1}{\z@}{-3.5ex plus -1ex minus
     -.2ex}{2.3ex plus .2ex}{\normalsize\bf}}
\def\subsection{\@startsection{subsection}{2}{\z@}{-3.25ex plus -1ex minus
 -.2ex}{1.5ex plus .2ex}{\normalsize\bf}}

 \def\thebibliography#1{\section*{References\markboth
  {REFERENCES}{REFERENCES}}\list
  {[\arabic{enumi}]}{\settowidth\labelwidth{[#1]}\leftmargin\labelwidth
  \advance\leftmargin\labelsep
  \usecounter{enumi}}
  \def\newblock{\hskip .11em plus .33em minus -.07em}
  \sloppy
  \sfcode`\.=1000\relax}
 
\catcode `\@=12

\makeatletter
\def\refptcirc{\lower\feyn@maxis \hbox to 0pt{\hss$\circ$\hss}}
\makeatother
\let\\\cr 

\begin{document}


\maketitle
\begin{abstract} {\small 
The spectrum of masses of the colorless states of the DW (De Donder-Weyl) Hamiltonian operator 
of quantum SU(2) Yang-Mills field theory on $\mathbb{R}^D$ 
obtained via the precanonical quantization is shown to be 
purely discrete and bounded from below. 
The scale of the mass gap is estimated to be of the order of magnitude of the 
scale of the ultra-violet parameter $\varkappa$ 
 introduced by precanonical quantization on dimensional grounds. }

\bigskip 

\end{abstract}

\vspace*{-142mm}
\hbox to 5.8truein{
Int.~J.~Geom.~Meth.~Mod.~Phys. {\bf 14} (2017) 1750123 \hfil 
\hbox to 1 truecm{ 
\hss \normalsize arXiv:1706.01766 [hep-th]} \vspace*{-4mm}
}
\hbox to 5.8truein{\hspace*{-4pt}\hrulefill}
\vspace*{135mm}


\newcommand{\beq}{\begin{equation}}
\newcommand{\eeq}{\end{equation}}
\newcommand{\beqa}{\begin{eqnarray}}
\newcommand{\eeqa}{\end{eqnarray}}
\newcommand{\nn}{\nonumber}

\newcommand{\half}{\frac{1}{2}}

\newcommand{\xt}{\tilde{X}}

\newcommand{\uind}[2]{^{#1_1 \, ... \, #1_{#2}} }
\newcommand{\lind}[2]{_{#1_1 \, ... \, #1_{#2}} }
\newcommand{\com}[2]{[#1,#2]_{-}} 
\newcommand{\acom}[2]{[#1,#2]_{+}} 
\newcommand{\compm}[2]{[#1,#2]_{\pm}}

\newcommand{\lie}[1]{\pounds_{#1}}
\newcommand{\co}{\circ}
\newcommand{\sgn}[1]{(-1)^{#1}}
\newcommand{\lbr}[2]{ [ \hspace*{-1.5pt} [ #1 , #2 ] \hspace*{-1.5pt} ] }
\newcommand{\lbrpm}[2]{ [ \hspace*{-1.5pt} [ #1 , #2 ] \hspace*{-1.5pt}
 ]_{\pm} }
\newcommand{\lbrp}[2]{ [ \hspace*{-1.5pt} [ #1 , #2 ] \hspace*{-1.5pt} ]_+ }
\newcommand{\lbrm}[2]{ [ \hspace*{-1.5pt} [ #1 , #2 ] \hspace*{-1.5pt} ]_- }

\newcommand{\pbr}[2]{ \{ \hspace*{-2.2pt} [ #1 , #2\hspace*{1.4 pt} ] 
\hspace*{-2.3pt} \} }
\newcommand{\nbr}[2]{ [ \hspace*{-1.5pt} [ #1 , #2 \hspace*{0.pt} ] 
\hspace*{-1.3pt} ] }

\newcommand{\we}{\wedge}
\newcommand{\nbrpq}[2]{\nbr{\xxi{#1}{1}}{\xxi{#2}{2}}}
\newcommand{\lieni}[2]{$\pounds$${}_{\stackrel{#1}{X}_{#2}}$  }

\newcommand{\rbox}[2]{\raisebox{#1}{#2}}
\newcommand{\xx}[1]{\raisebox{1pt}{$\stackrel{#1}{X}$}}
\newcommand{\xxi}[2]{\raisebox{1pt}{$\stackrel{#1}{X}$$_{#2}$}}
\newcommand{\ff}[1]{\raisebox{1pt}{$\stackrel{#1}{F}$}}
\newcommand{\dd}[1]{\raisebox{1pt}{$\stackrel{#1}{D}$}}
\newcommand{\der}{\partial}

\newcommand{\Om}{\Omega}
\newcommand{\om}{\omega}
\newcommand{\eps}{\epsilon}
\newcommand{\si}{\sigma}
\newcommand{\Lm}{\bigwedge^*}

\newcommand{\inn}{\hspace*{2pt}\raisebox{-1pt}{\rule{6pt}{.3pt}\hspace*
{0pt}\rule{.3pt}{8pt}\hspace*{3pt}}}
\newcommand{\sro}{Schr\"{o}dinger\ }
\newcommand{\bm}{\boldmath}
\newcommand{\vol}{\omega}
               \newcommand{\dvol}[1]{\der_{#1}\inn \vol}

\newcommand{\bd}{\mbox{\bf d}}
\newcommand{\bder}{\mbox{\bm $\der$}}
\newcommand{\bI}{\mbox{\bm $I$}}

\newcommand{\be}{\beta} 
\newcommand{\ga}{\gamma} 
\newcommand{\de}{\delta} 
\newcommand{\Ga}{\Gamma} 
\newcommand{\gmu}{\gamma^\mu}
\newcommand{\gnu}{\gamma^\nu}
 \newcommand{\ka}{\varkappa} 
 \newcommand{\la}{\lambda}
\newcommand{\hka}{\hbar \kappa}
\newcommand{\al}{\alpha}
\newcommand{\lapl}{\bigtriangleup}
\newcommand{\psib}{\overline{\psi}}
\newcommand{\Psib}{\overline{\Psi}}
\newcommand{\Phib}{\overline{\Phi}}
\newcommand{\derts}{\stackrel{\leftrightarrow}{\der}}
\newcommand{\what}[1]{\widehat{#1}}

\newcommand{\bx}{{\bf x}}
\newcommand{\bk}{{\bf k}}
\newcommand{\bq}{{\bf q}}

\newcommand{\omk}{\omega_{\bf k}} 
\newcommand{\lpl}{\ell}
\newcommand{\zb}{\overline{z}} 

\newcommand{\deltab}{\bar \delta}

\newcommand{\gammas}{\stackrel{\raisebox{-7pt}{\tiny$\,*$}}{\gamma}}

\newcommand{\dv}{\mbox{\sf d}}

\newcommand{\BPsi}{\mathbf{\Psi}} 
\newcommand{\BPhi}{\mathbf{\Phi}}
\newcommand{\BH}{{\bf H}} 
\newcommand{\BS}{{\bf S}} 
\newcommand{\BN}{{\bf N}} 

\newcommand{\rd}{\mathrm{d}}
\newcommand{\ri}{\mathrm{i}}
\newcommand{\Tr}{\mathrm{Tr}} 
\newcommand{\re}{\mathbbl{e}}   
\newcommand{\Ai}{\mathtt{A\hspace{-1pt}i}} 
\newcommand{\Bi}{\mathtt{B\hspace{-1pt}i}}

\newcommand{\dslash}{d \hspace{-0.7ex}\rule[1.3ex]{0.8ex}{.06ex}} 
\newcommand{\kslash}{k \hspace{-1.0ex}\rule[1.3ex]{0.8ex}{.06ex}} 


\section{Introduction}

In this paper, we study the spectrum of the DW (De Donder-Weyl \cite{dw,kastrup}) 
Hamiltonian operator  
which was obtained within the precanonical quantization of pure 
YM theory  \cite{my-ym2003,my-ehrenfest}: 
\beq \label{dwhop}
\what{H} = 
 \frac{1}{2} \hbar^2\varkappa^2 \frac{\der^2}{\der A_a^\mu\der A^a_\mu } 
- \frac{1}{2}\ri g\hbar\varkappa  C^a{}_{bc}A^b_\mu A^c_\nu 
\gamma^\nu \frac{\der}{\der A^a_\mu } \; . 
\eeq
Here and in what follows,   $A_a^\mu$ are YM gauge potentials, $\gamma^\mu$ are Dirac matrices, 
$C^{a}{}_{bc}$ are the structure constants of the gauge group, 
$g$ is the gauge coupling constant, 
and $\ka$ is an ultraviolet  parameter of 
 dimension $\mathtt{length}^{-(n-1)}$  in $n$ space-time dimensions,  
which 
is introduced by precanonical quantization (when the differential forms 
corresponding to dynamical variables are represented by Clifford-algebra-valued 
operators \cite{my-quant}). 
In what follows,   we also denote $\frac{\der}{\der A^a_\mu }$ as ${\der}_{A^a_\mu }$ 
and $\frac{\der}{\der x^\mu }$ as $\der_\mu$.

Let us recall that the importance of the DW Hamiltonian operator within the precanonical 
quantization of fields is that 
the evolution of Clifford-algebra-valued 
precanonical wave functions $\Psi(A,x)$ on the bundle of field variables $A$ over 
 space-time (whose coordinates are $x^\mu$) is controlled by the Dirac-like equation 
\cite{my-quant} 
\beq \label{schreq}
\ri\hbar\ka\gamma^\mu\der_\mu \Psi = \what{H}\Psi \,.
\eeq 
This equation appears as a generalization of the Schr\"odinger equation to field theory 
when all space-time variables are being treated 
 on an equal footing. In this formulation, the only entities which propagate 
 are different modes of precanonical wave function on the total space of field 
 variables and space-time variables. 
Correspondingly, the masses of propagating modes of (the precanonical wave function of) 
quantum fields are given by the spectrum of $\frac{1}{\ka}\what{H}$.  
Note that in the classical limit (\ref{schreq}) is  consistent with the classical field equations: 
in  \cite{my-ehrenfest},  we have shown that the classical YM field equations arise from  precanonical formulation as the equations satisfied by the expectation values of the corresponding 
field operators defined by precanonical quantization.  
The expectation values  are calculated using the 
 scalar product 
 \beq  \label{sp}
 \langle \Psi|\Psi\rangle = \Tr \int\! [\rd A] \Psib \Psi , 
 \eeq
where $[\rd A]:=\prod_{\mu,a} \rd A^a_\mu$ and $\Psib:=\gamma^0\Psi^\dagger\gamma^0$. 
The derivation of this generalization of the  Ehrenfest theorem  requires that 
$\what{H}$ is pseudo-Hermitian with respect to (\ref{sp}): $\overline{\what{H}} = \what{H}$ 
(cf. \cite{mostafa}). 

A relation between the precanonical field quantization 
and the canonical quantization in  
the functional Schr\"odin\-ger representation \cite{hatfield} 
was established in  \cite{my-schrod} and extended to quantum YM fields 
in \cite{my-ym2016}. In \cite{my-schrod},  it was argued that the standard 
QFT in the functional Schr\"odin\-ger representation 
can be derived 
from precanonical formulation in the limiting case of 
an 
infinitely large 
ultra-violet scale $\varkappa$. 
  In this limiting  case, the Schr\"odinger wave functional of a field configuration 
  $y=y(\bx)$ at a moment of time $t$ 
  is expressed as a 
  product integral \cite{prodint} of precanonical wave functions restricted  
  to this field configuration.  The canonical functional derivative 
    Schr\"odin\-ger equation can be derived from (\ref{schreq}) by means of  
  the space+time splitting $x^\mu \rightarrow (t, \bx)$ and restriction to the 
  surface representing the abovementioned field configuration.
  
This result is consistent with the fact that the standard formulation 
of QFT leads to the meaningful results only 
 upon regularization and renormalization. 
Precanonical quantization with its inbuilt ultraviolet scale $\ka$ appears as 
an ``already regularized" QFT because the divergent integrals in the usual formulations 
are replaced  
by expressions containing $\ka$ in the precanonical formulation \cite{my-tobe}.
However, 
unlike the nonlocal theories or theories on noncommutative,  discrete or fractal space-times, or theories which attempt to take into account the effects of quantum gravity, 
 LQG or string theories, 
in precanonical quantization,  the finite expressions 
are obtained without manually modifying the relativistic 
space-time at small distances in order to regularize the infinities.   
The quantum-gravitational geometry of space-time has been discussed within the precanonical quantization of Einstein  gravity 
 in \cite{my-qgr,my-metric}. 

 Our interest in the study of the spectrum of (\ref{dwhop}) originates from the observation 
\cite{my-quant} 
 that the spectrum of the DW Hamiltonian of a free scalar field is that of the harmonic oscillator in the field space: 
 $\ka m (N+1/2)$, where $N \!\in\! \mathbb{Z_+}$ is a non-negative integer.  
 When perturbed, this system radiates excitations corresponding to the 
 only allowed transitions between the nearby levels with $\Delta N = \pm 1$, that corresponds 
 to the excitations of mass $m$, which we usually interpret as 
 the 
 free massive quantum scalar particles. 
 In the case of free massless fields, the spectrum of the DW Hamiltonian is continuous and the 
 corresponding propagating excitations are massless. Hence, the discreteness or continuity 
 of the spectrum of (\ref{dwhop})  
  will tell us whether  the propagating excitations of quantum pure YM 
 field are massive or massless. 
 
 
 Let us recall that precanonical quantization of fields \cite{my-quant} 
 is the result of quantization 
 of the Poisson-Gerstenhaber (PG) brackets found in \cite{my-pbr} within 
 the DW Hamiltonian theory \cite{dw,kastrup}.  Those brackets are defined on differential forms 
 which represent the 
 dynamical variables or observables of the theory. Their generalization to singular DW Hamiltonian theories is discussed in \cite{my-dirac}. The Clifford-Dirac algebra appears 
 as the result of quantization of bi-graded PG brackets of forms \cite{my-quant,geom-q}. 
 The Dirac-like precanonical analogue of the Schr\"odinger equation in (\ref{schreq}) 
 is a quantum counterpart of the classical expression of the DW Hamiltonian equations 
 in terms of the PG brackets, where the bracket with the DW Hamiltonian function 
 $H$ is related to the operation of the exterior differential of forms 
 \cite{my-pbr}, whose quantum version is the Dirac operator \cite{my-quant,geom-q}.

\section{Elements of precanonical quantum YM  on $\mathbb{R}^\mathrm{D}$}

\newcommand{\bA}{\mathbf{A}}
\newcommand{\bB}{\mathbf{B}}
\newcommand{\bC}{\mathbf{C}}
\newcommand{\bD}{\mathbf{D}}
\newcommand{\bX}{\mathbf{X}}
\newcommand{\bY}{\mathbf{Y}}
\newcommand{\bF}{\mathbf{F}}
\newcommand{\bgamma}{\boldsymbol\gamma} 

In this paper, we consider a Euclidean theory over $\mathbb{R}^\mathrm{D}$.  
The DW Hamiltonian operator is given by 
\beq \label{dwh-eigenv}
\frac1\varkappa \what{H} = 
-  \frac{1}{2} \hbar^2\varkappa \der^2_ {A^a_i A^a_i} 
- \frac{1}{2}\ri g\hbar  C^a{}_{bc}A^b_i A^c_j 
\gamma^j \der_{A^a_i }  , 
\eeq 
where $\gamma^i$, $i=1,...,D$,  are anti-Hermitian and $\gamma^i\gamma^j+\gamma^j\gamma^i
= - \delta^{ij}$. This operator acts on the space of Clifford-valued wave 
functions\footnote{For the relevant mathematics consult e.g. \cite{hestenes,pavsic,habetha}.} 
\beq
\Psi (A) = \psi(A)+\psi_i(A)\gamma^i+\psi_{i_1i_2}(A)\gamma^{i_1i_2}+...+ 
\psi_{i_1i_2...i_D}(A)\gamma^{i_1i_2...i_D} \,.
\eeq 
It is Hermitian on the space of those functions  equipped with the scalar 
product 
\beq \label{scpr1}
\left< \Phi,\Psi \right> = 
\int\![\rd A]\ [ \Phi^{*{\sf r}} \Psi ]_0 , 
\eeq
where ${}*$ is the complex conjugation of the components of $\Phi$, 
${}^{\sf r}$ is the reversal anti-automorphism on the Clifford algebra such that, e.g.,  
$(\gamma^i\gamma^j)^{\sf r}= \gamma^j\gamma^i$, $[...]_0$ denotes the scalar part of the 
Clifford number under the square brackets, and $[\rd A] :=\prod_{i a} \rd A_i^a$. 
This product is not positive definite, however, because 
\beq
\Psi^{*{\sf r}} = \psi^*(A)+\psi^*_i\gamma^i - \psi^*_{i_1i_2}(A)\gamma^{i_1i_2}+...+ 
(-1)^{D(D-1)/2} \psi^*_{i_1i_2...i_D}(A)\gamma^{i_1i_2...i_D} \,.
\eeq 
The positive definite scalar product is given by 
\beq \label{scpr2}
\left<\!\left<\Phi,\Psi \right>\!\right> = 
\int\![\rd A]\ [\Phi^{*{\sf r}{\sf a}} \Psi ]_0 \,,
\eeq
where ${}^{\sf a}$ is the main Clifford automorphism: $\gamma^i{}^{\sf a} = - \gamma^i$. 
$\what{H}$ is pseudo-Hermitian with respect to  (\ref{scpr2}) in the sense that 
$\left<\!\left<\Phi, \right.\right. \what{H}\!\left.\left.\Psi \right>\!\right> = 
\left<\!\left<\right.\right.\!\what{H}{}^{\sf a}\Phi,\left.\left. \Psi \right>\!\right>$. 
The Hermitian conjugate matrix $\Psi^\dagger =  \Psi^{*{\sf r}{\sf a}}$, so that 
the integrand in (\ref{scpr2}) 
is just the Frobenius inner product of matrices. 
Note that, in pseudo-Euclidean space-times, 
the integrand in 
(\ref{scpr1}) coincides with $[\beta\Phi^\dagger\beta\Psi]_0=:[\Phib\Psi]_0$ used in our previous papers. 

 \section{
 Spectrum of DW Hamiltonian of SU(2) YM}

We are interested in the eigenvalue problem for the DW Hamiltonian operator devided by $\ka$:
\beq \label{dwhmu}
\frac{1}{\ka}\what{H}\Psi = \mu\Psi . 
\eeq
In the case of SU(2) gauge group,  the structure constants $C^a{}_{bc}$  are the Levi-Civita symbol $\epsilon_{abc}$ and $A^a_i$ is a triplet of vector 
fields $\bA, \bB, \bC$ with the components $A_i,B_i,C_i$, $i=1,...,D$. In this notation, 
the operator  
in Eq. (\ref{dwh-eigenv}) takes the form 
\beq \label{dwh3} 
\frac{1}{\ka}\hat{H}=
-\frac{\hbar^2\ka}{2} \big(\der_{\bA\bA} + \der_{\bB\bB} + \der_{\bC\bC}\big) 
 + \frac{g\hbar}{2} \left( \slashed \bA \hat{L}_{\bB\bC} 
 + \slashed \bB \hat{L}_{\bC\bA} 
 + \slashed \bC  \hat{L}_{\bA\bB} \right) 
 \,,
\eeq
where $\der_{\bA\bA} := \sum_i \der_{A_iA_i}$, $\slashed \bA:= A_i\gamma^i$ and  
$\hat{L}_{\bA\bB} := \ri( \bA\cdot\der_\bB - \bB\cdot \der_\bA) 
= \sum_i \ri (A_i\der_{B_i} - B_i\der_{A_i})$. 
 Let us notice the permutation symmetry between $\bA$, $\bB$ and $\bC$ 
 which will allow us to simplify the problem. 

In order to 
  analyze the spectrum of (\ref{dwh3}),  let us rewrite  
it in the form 
\begin{align} \label{est00}
\frac{1}{\ka}\what{H} = 
&-\frac{\hbar^2\ka}{2}\der_{\bA\bA}  
+ \frac{\ri g\hbar}{2}\Big( \big( \slashed \bB \bC - \slashed \bC \bB \big)\cdot \der_\bA
 + 
\bA\cdot\big (\slashed \bC\der_{\bB} - \slashed \bB\der_{\bC}\big) 
  \Big) 
 + \frac{g\hbar}{2}  \slashed \bA \hat{L}_{\bB\bC} 
 \\ 
&-\frac{\hbar^2\ka}{2} \big(\der_{\bB\bB}  + \der_{\bC\bC}\big) 
 =: \what{G}_\bA  -\frac{\hbar^2\ka}{2} \big(\der_{\bB\bB}  + 
\der_{\bC\bC}\big)  \,. \nn
\end{align} 
Here,  the permutation symmetry between  $\bA$, $\bB$ and $\bC$ is not manifest anymore. 
In fact, the arguments below can be 
 presented in a manifestly  permutation symmetric way 
by considering $\frac{1}{\ka}\what{H}$ in the form 
\beq \label{h3}
\frac13 \left( \what{G}_\bA +\what{G}_\bB + \what{G}_\bC  
- \hbar^2\ka\big(\der_{\bA\bA}  + \der_{\bB\bB}  + \der_{\bC\bC}
\big) \right) \,,
\eeq 
where $\what{G}_\bB$ and  $\what{G}_\bC$ 
are obtained from $\what{G}_\bA$ by 
cyclic permutation of $\bA, \bB, \bC$. 
However, it would make the formulae more cumbersome and 
the presentation lengthier 
without adding to its essence.

Thus, we proceed by  
 concentrating on the operator $\what{G}_\bA$ which has absorbed 
all the terms with $\bA$ and $\der_\bA$ in (\ref{est00}).   
At first we note that 
$\hat{L}_{\bB\bC} = \sum_i \ri (B_i\der_{C_i} - C_i\der_{B_i})$ 
is similar to  the 
angular momentum operator $\hat{L}_{z}$ in quantum mechanics.  
Therefore, the eigenvalues of $\hat{L}_{\bB\bC}$ are 
 integers \ 
 $
 {m} := \sum_i m_i = m_1 + m_2+...+m_D \in \mathbb{Z}$ 
 with the admissible  values of each $m_i$ being limited by the  
 quantum numbers ${l}_i$ of the 
 operators $\what{L}_{A_iB_i}^2 + \what{L}_{A_iC_i}^2 + \what{L}_{B_iC_i}^2$ 
 for the corresponding value of $i$ (no summation over $i$ here): $|m_i|\leq l_i$.  
Similarly, the eigenvalues of $\hat{L}_{ij}:=\ri (B_i\der_{C_j}- C_i\der_{B_j})$ 
are integers $m_{ij} \in \mathbb Z$ whose absolute values are restricted 
by the quantum numbers $l_{ij}$ of the operators $\hat{L}_{ij}^2$.   
 Because $[\hat{L}_{\bB\bC}, \hat{L}_{ij}] =0$,  
we can chose the basis $|m,m_{ij}\rangle$ in which both operators are diagonal.
%
%
 Then the operator in the first line of (\ref{est00}) takes the form 
\beq \label{est01}
\what{G}_\bA:= -\frac{\hbar^2\ka}{2}\der_{\bA\bA}  
+ \frac{i g\hbar}{2} \big( \slashed \bB \bC - \slashed \bC \bB \big)\cdot \der_\bA
+ \frac{g\hbar}{2} \big( 
 {m}\slashed \bA 
 +  m_{ij} A_i\gamma^j \big)  \,.
\eeq

 Let us estimate the spectrum of $\what{G}_\bA$   
 when $\bB$ and $\bC$ are treated 
as external parameters. By denoting 
$\widetilde{m}_{ij}:= {m} \delta_{ij} + m_{ij}$  
and 
 $\bD^2 := \bB^2\bC^2 - (\bB\cdot\bC)^2$ 
we  rewrite (\ref{est01}) in the form 
\beq \label{est02a}
\what{G}_\bA = \frac{\hbar^2\ka}{2}\left(
\ri\der_{A_i} + \frac{g}{2\hbar\ka}\gamma^j(B_jC_i-C_jB_i)\right)^2 
 + \frac{\,\,g^2}{4 \ka}\bD^2  + \frac{g\hbar}{2}\gamma^i 
 \widetilde{m}_{ij} A_j  \,,
\eeq 
which resembles the  magnetic Schr\"odinger 
operator  in $A$-space with 
the Clifford-algebra-valued analogs of electric and magnetic potentials in that space.  
The ``magnetic potential" term $\mathcal{A}_i:=\frac{g}{2\hbar\ka}\gamma^j(B_jC_i-C_jB_i)$ 
is  
constant in $A$-space: $\der_\bA\mathcal{A} =0$, 
hence it 
can be expected to contribute 
only a phase factor to the solution and does not contribute to the 
eigenvalues of $\what{G}_\bA$.  
The  ``electric potential" term contains the linear term in $A_i$
and the constant term 
 $\frac{\,\,g^2}{4\ka}\bD^2$ (in the sense that $\der_\bA\bD^2 =0$) 
 which just shifts the eigenvalues.

 At first, let us consider the case when all quantum numbers $\widetilde{m}_{ij} = 0$.  
In order to find the magnetic phase factor $U$, let us set $\Psi (A) = U(A;B,C) \Phi(A)$, 
so that all the parametric dependence of the eigenstates of $\what{G}_\bA$ of $\bB$ and $\bC$ 
is absorbed in the phase factor $U$. Obviously, the eigenvalue problem 
\begin{align}  \label{036}
\begin{split}
  \left( -\frac{\hbar^2\ka}{2}\der_{\bA\bA}  
+ \frac{i g\hbar}{2} \big( \slashed \bB \bC - \slashed \bC \bB \big)\cdot \der_\bA \right)\Psi 
= \chi \Psi 
\end{split} 
\end{align}
leads to the equations 
\begin{align}
-{\hbar^2\ka} \der_\bA U + \frac{i g\hbar}{2} \big( \slashed \bB \bC - \slashed \bC \bB \big) U 
&= 0 \label{037}\\ 
-\frac{\hbar^2\ka}{2}\der_{\bA\bA} U
+ \frac{i g\hbar}{2} \big( \slashed \bB \bC - \slashed \bC \bB \big)\cdot \der_\bA  U 
&= \xi U \label{038}\\ 
 -\frac{\hbar^2\ka}{2}\der_{\bA\bA} \Phi = (\chi -\xi) \Phi \,.  \label{039}
\end{align}
From (\ref{037}) it follows 
\beq
-\frac{\hbar^2\ka}{2}\der_{\bA\bA} U + \frac{g^2}{4\ka} \bD^2 U = 0 .  
\eeq
By substituting (\ref{037}) in (\ref{038}), we obtain 
\beq
-\frac{\hbar^2\ka}{2}\der_{\bA\bA} U +  \frac{g^2}{2\ka} \bD^2 U = \xi U  . 
\eeq
Therefore, $\xi = \frac{g^2}{4\ka} \bD^2$ and 		eq. (\ref{036}) for $\Psi$ 
is equivalent to the following equations for $\Phi(A)$ 
\beq \label{0312}
-\frac{\hbar^2\ka}{2}\der_{\bA\bA}\Phi + \frac{g^2}{4\ka} \bD^2 \Phi = \chi \Phi 
\eeq 
and the phase factor $U$ 
\beq \label{0313}
\gamma^i\der_{A_i} U = - \frac{\ri g}{2\hbar\ka} \gamma^{ij}(B_iC_j-B_jC_i) U . 
\eeq 
The latter is obtained by contracting (\ref{037}), which is not an integrable equation,  
with $\gamma^i$.  The solutions of (\ref{0313}) which satisfy the condition $U(A\!=\!0)=1$ 
can be viewed as a hypercomplex generalizations of the exponential function (different from 
those considered in Clifford analysis (c.f. \cite{habetha,sommen81,malonek})).  


 From (\ref{0312}) we conclude that  $\chi \ge \frac{g^2}{4\ka}\bD^2$. 
 Hence, for the states of (\ref{est02a}) corresponding to the lowest 
 quantum numbers 
 $\tilde{m}_{ij}=0$,   
 we can  estimate the lower bound of the 
 eigenvalues of $\what{G}_\bA$: 
 \beq \label{estG1}
 \what{G}_\bA \geq \frac{\,\,g^2}{4 \ka}\bD^2 \,.
 \eeq  
 This estimation is suitable for the eigenvalues of $\what{H}$ 
 corresponding to the``colorless" states (i.e. those invariant with respect to the 
 internal SU(2) rotations).  
 
The states with non-vanishing quantum numbers $m$ and/or $m_{ij}$  correspond  
to the ``colored" states. Let us consider the influence of the last term in  
(\ref{est02a}) on the estimation (\ref{estG1}).  To this end,  we need to study 
the eigenvalue problem of $\what{G}_\bA$ in more detail. 
  
\newcommand{\bt}{\mathbf{t}}  
\newcommand{\mm}{\mathbf{m}}
\newcommand{\bp}{\mathbf{p}}
\newcommand{\bc}{\mathbf{c}} 
\newcommand{\uj}{\underline{j}} 
\newcommand{\ui}{\underline{i}} 
\newcommand{\uk}{\underline{k}} 
  
We use the fact that the second term in (\ref{est02a}) leads to the phase factor 
$U$ in the wave function, which transforms (\ref{036}) to (\ref{0312}). 
Let us omit  in (\ref{est02a}) the unnecessary coefficients and the additive term 
$\bD^2$, 
 and consider the eigenvalue problem 
\beq \label{eig0}
\left( - \der_{\bA\bA} + g \gamma^i\widetilde{m}_{ij}A_j \right)\Phi(\bA)=\lambda'\Phi (\bA) \,
\eeq
for arbitrary fixed values of $\widetilde{m}_{ij}=m\delta_{ij}+m_{[ij]}$. 
The estimation (\ref{estG1}) will hold also for the ``colored" states 
if $\lambda'\geq 0$ for any $\widetilde{m}_{ij}$. 

To check if this is the case,  let us first consider a special case when $\widetilde{m}_{ij}$ has only one 
independent non-vanishing component $m_{12}$, which can be either positive or 
negative integer.  Then (\ref{eig0}) reads 
\beq \label{eig12}
\left( - \sum_{k\neq 1,2}\der_{A_kA_k} 
- \der_{A_1A_1}  - \der_{A_2A_2} + gm_{12}(\gamma^1A_2 -\gamma^2A_1) 
\right) \Phi = \lambda'\Phi \,.
\eeq
By separating the variables with $k\neq 1,2$ and $k = 1,2$, 
 we obtain a two-dimensional eigenvalue problem 
\beq \label{eig12aa}
\left( - \der_{A_1A_1}  - \der_{A_2A_2} + gm_{12}(\gamma^1 A_2 -\gamma^2A_1)\right) \Phi(A) = \lambda''\Phi(A) , 
\eeq
which can be seen as a two-dimensional Clifford-algebraic generalization of the Airy equation 
[22, 24] for the Clifford-valued wave function  
\beq 
\Phi (A)= \phi(A) + \phi_i(A)\gamma^i. 
\eeq 
Using the notation $\Delta_{12}:=\der_{A_1A_1}  + \der_{A_2A_2}$, let us write (\ref{eig12aa}) 
in the component form 
\begin{align} \label{eig12a}
 \begin{split}
-\Delta_{12}\phi - gm_{12}(A_2\phi_1-A_1\phi_2) &=\lambda''\phi
\\ 
-\Delta_{12}\phi_1 + gm_{12}A_2\phi &= \lambda''\phi_1
\\
-\Delta_{12}\phi_2 - gm_{12}A_1\phi &= \lambda''\phi_2 
\\
A_1\phi_1 + A_2\phi_2&= 0 \\
A_i\phi_k &= 0 \quad \mathrm{for}\; k\neq 1,2 .
\end{split}
 \end{align}
The solutions are $\phi_1 = \frac{\pm A_2}{|A|}\phi$, $\phi_2 = \frac{\mp A_1}{|A|}\phi$ 
and $\phi_k = 0$ for $k\neq 1,2,$ 
where $|A|:=\sqrt{A_1^2+A_2^2}$, and the non-vanishing components 
of $\Phi$ obey 
\begin{align} \label{eig12f}
-\Delta_{12}\phi\, \mp\, gm_{12}|A| \phi\, &=\lambda'' \phi  \\
-\Delta_{12}\phi_i \pm gm_{12}|A| \phi_i &=\lambda'' \phi_i, \quad i=1,2.  
\end{align} This system has a discrete positive spectrum both for positive and negative 
values of $m_{12}$. For example, with the upper choice of the sign, the eigenfunctions $  \Phi(u)$ have a vanishing scalar part $\phi$ for $m_{12}>0$ or vanishing vector components $\phi_{1,2}$ for $m_{12}<0$. 
In both cases, the separation of angular and radial variables in the two-dimensional Laplacian 
$\Delta_{12}$ leads to the radial equation for the lowest (vanishing) quantum number of the orbital angular momentum corresponding to the rotations in $(A_1,A_2)$-plane: 
\beq
\left(-\der_{\rho\rho} - \frac{1}{\rho}\der_\rho + g |m_{12}|\rho \right)\phi (\rho)=\lambda'' \phi(\rho),  
\eeq
where $\rho := |A| \geq 0$. The linear growth of the potential term ensures that the 
spectrum is discrete with the eigenvalues 
corresponding to the boundary condition $\der_A\left<\Phi^\dagger\Phi\right>=0$ at $|A|=0$.  
%
%
Hence,  for any value of $m_{12}$ the spectrum of (\ref{eig12aa}) is positive and discrete.  
Correspondingly, the spectrum of (\ref{eig12}) is positive and bounded from below by the lowest eigenvalue of (\ref{eig12aa}): $\lambda' = \lambda'' + (p_k)^2$, where the last term $(p_k)^2>0$ is the continuous 
spectrum of $- \sum_{k\neq 1,2}\der_{A_kA_k}$.

\newcommand{\rank}{\mathsf{rank}}

Similarly, for arbitrary $\widetilde{m}_{ij}$, we obtain from (\ref{eig0}) 
\begin{align} \label{eig12b}
 \begin{split}
 - \der_{\bA\bA}\phi - g \widetilde{m}_{ij}A_j\phi_i &=\lambda''\phi \\
- \der_{\bA\bA}\phi_i  + g \widetilde{m}_{ij}A_j\phi&=\lambda''\phi_i \\
 \widetilde{m}_{[ij}A_j\phi_{k]}&=0.
\end{split}
 \end{align} 
The last equation is solved by $\phi_i = \pm \frac{\widetilde{m}_{ij}A_j}{|\widetilde{m}A|} \phi$, where 
 $|\widetilde{m}A| := \sqrt{\widetilde{m}_{ij}A_j\widetilde{m}_{ik}A_k}$. Then (\ref{eig12b}) yields
 \begin{align} \label{eig12cd}
 - \der_{\bA\bA}\phi\, & \mp g  |\widetilde{m}A|\phi\ =\lambda''\phi ,\\
 - \der_{\bA\bA}\phi_i &\pm g |\widetilde{m}A|\phi_i =\lambda''\phi_i .
 \end{align} 
The positive discrete spectrum is obtained 
either for purely scalar wave functions $\Phi = \phi$ or purely vector ones $\Phi = \phi_i\gamma^i$ 
for the lower or upper choice of the signs, respectively. 
The eigensolutions correspond to the boundary 
conditions $\der_\bA \left<\Phi^\dagger\Phi\right>=0$ at $|\widetilde{m}A|=0$. 
If $\rank(\widetilde{m}_{ij}) <D$, the spectrum has the form $\lambda'' = \lambda' + p^2$, where $\lambda'$ are purely discrete and $p^2$ is the continuous norm  squared of a vector in  the Euclidean $(D- \rank(\widetilde{m}))$-dimensional space.

 \newcommand{\bN}{\mathbf{N}}

 \newtheorem{lma}{Lemma} 
  
Therefore, it is  shown that 
\begin{lma}  
For any non-vanishing 
values of quantum numbers 
$m$ and $m_{ij}$
the spectrum of (\ref{eig0}) 
is discrete and bounded from below.  
\end{lma} 

Note that this Lemma is the necessary condition for the validity of the estimation (\ref{estG1}) for the ``colored" states with $\widetilde{m}_{ij} \neq 0$. The sufficient 
condition has to take into account the non-commutativity of the Clifford-algebra-valued potential term $g\gamma^i\widetilde{m}_{ij}A_j$ with the phase factor $U$ which is a 
solution of (\ref{0313}). It requires an analysis of a generalization of (\ref{eig0}) 
with the potential term $g U \gamma^i\widetilde{m}_{ij}A_j U^{-1}$ which will parametrically 
depend on $\bB$ and $\bC$. As our main interest here is the spectrum of the ``colorless" 
states of $\what{H}$ with vanishing $\widetilde{m}_{ij}$, we 
leave this part of the problem beyond the scope of the paper.

Now, using this result in eq. (\ref{est00}), we can write the following inequality 
for the operator $\frac{1}{\ka}\what{H}$,  
which is valid at least for the colorless states:  
\beq \label{est02}
\frac{1}{\ka}\what{H}\geq  -\frac{\hbar^2\ka}{2} \big(\der_{\bB\bB} +\der_{\bC\bC}\big) 
 +  \frac{ g^2}{4\ka} \big(\bB^2\bC^2 - (\bB\cdot\bC)^2 \big) \,. 
\eeq
The spectrum of the operator in the r.h.s. of (\ref{est02}) is purely discrete 
and  bounded from  below, as it was already proven e.g. in sect. 7 of \cite{simon83} 
using the Fefferman-Phong theorem. Then the inequality (\ref{est02}) 
proves the main assertion of this paper: 
\newtheorem{thm}{Theorem} 
\begin{thm} 
The spectrum of  colorless states of the DW Hamiltonian operator of quantum SU(2) Yang-Mills field 
is purely 
discrete and bounded from below. 
\end{thm}
An immediate consequence of the discreteness of the spectrum of colorless states 
of the DW Hamiltonian operator is the nonvanishing gap between the ground state and the lowest colorless excited state. It means that the first  propagating colorless excited mode of quantum YM field is massive. 
 
 \medskip 
 
 It is interesting to consider the eigenvalue problem of the operator in the r.h.s. of (\ref{est02}) in more detail. 
In order to estimate 
 its spectrum,  let us simplify the problem by approximating the potential term 
 $\bB^2\bC^2 - (\bB\cdot\bC)^2 = \bB^2\bC^2 \sin^2 \theta$, where $\theta$ 
 is the angle between $\bB$ and $\bC$,  by  its average over $\theta\in[0,2\pi]$: 
\beq
\frac{1}{2\pi}\int_0^{2\pi} \!\rd \theta\,  \bB^2\bC^2 \sin^2 \theta = 
\frac{1}{2} \bB^2\bC^2 \,.
\nn
\eeq 
 Then the estimation in (\ref{est02}) is replaced by 
 an approximate inequality 
 \beq \label{est02ap}
\frac{1}{\ka}\what{H} 
\gtrsim 
-\frac{\hbar^2\ka}{2} \big(\der_{\bB\bB} +\der_{\bC\bC}\big) 
 +  \frac{ g^2}{8\ka} \bB^2\bC^2  \,. 
\eeq
By employing the argument by B. Simon in sect.  2 of \cite{simon83}
and using the exactly known ground state of isotropic harmonic 
oscillator in $D$-dimensions,  we further obtain 
\beq \label{est4}
\frac{1}{\ka}\what{H} \gtrsim 
-\frac{\hbar^2\ka}{2} \der_{\mathbf{B}\mathbf{B}} 
  + \frac{Dg\hbar }{4} |\mathbf{B}| \,.
\eeq 
By going to the spherical coordinates in $\mathbf{B}$-space 
 and neglecting the knowingly positive contribution of the square of the orbital  
 angular momentum operator in $\bB$-space, we obtain a further approximate estimation 
\beq \label{estR}
\frac{1}{\ka}\what{H}  \gtrsim 
- \frac{\hbar^2\ka}{2} \left(\der_{rr} + \frac{D-1}{r}\der_r \right) 
+ \frac{Dg\hbar }{4} r 
=: \what{R} 
 \,, 
\eeq 
where $r$ denotes $|\mathbf{B}|$. 
 The eigenvalue problem $\what{R} f = \mu f$ with $r\in \mathbb{R}_+ =[0,\infty)$ 
leads to the solutions which interpolate between the Bessel function 
$J_{\frac{D}{2}-1}(\sqrt{\mu}r)/{r^{\frac{D}{2}-1}} $ at $r\rightarrow 0$ and the 
Airy function $\mathrm{Ai}(a (r - b)) $ with $a = \left( \frac{Dg}{2\hbar\ka}\right)^{1/3}$ 
and $b=\frac{4\mu}{Dg\hbar} $    at $r\rightarrow \infty$. 
The spectrum is discrete and positive, and it follows from the condition 
$\der_r f^2=0$ at $r=0$: $\mu_K = f(K,D) \left( \frac{g^2D^2\hbar^4\ka}{32}\right)^{1/3}$, 
where  the coefficients $f(K,D)$ labeled by $K\in \mathbb{Z}_+$ 
are given by the roots of $f(r)$ and its derivatives (similarly to the one dimensional 
$|x|$ problem in quantum mechanics, see \cite{bouncer,airy-book,hohlfeld}).

At $D=3$,  the spectral problem for the operator  in the r.h.s. of (\ref{estR}) 
is  quasi-exactly solvable \cite{ushveridze,turkish}. Namely,  an exact solution can be 
obtained for the lowest eigenstates  corresponding to the vanishing orbital angular momentum quantum number in $\bB$-space. 
This fact allows  us to estimate the lower bound of 
$\frac{1}{\ka}\what{H} $ in the physically relevant case of $D=3$ which 
can be related  to the  YM theory in $(3+1)$-dimensional  pseudoeuclidean space-time 
in the temporal gauge $A^a_0 =0$ (c.f. \cite{my-ym2003}).    
In this case, 
\beq
\frac{1}{\ka}\what{H} \gtrsim  \what{R} = 
- \frac{\hbar^2\ka}{2} \left(\der_{rr} + \frac{2}{r}\der_r \right) 
+ \frac{3}{4}g\hbar  r \,.
\eeq 
The solutions of the eigenvalue problem $\what{R} f = \mu f$ 
 (up to a normalization factor $\mathcal{N}$) are 
\beq 
f(r) = \mathcal{N} r^{-1}\mathtt{Ai} \left( \gamma r - \delta\right) \,,
\eeq  
 where 
 \beq
 \gamma = \left(\frac{3 g }{2\hbar\ka}\right)^{1/3}, 
 \quad 
 \delta = \mu \left(\frac{32}{9 g^2\hbar^4\ka} \right)^{1/3} \,.
 \eeq
Note that the wave functions $f(r)$ are  normalized to a delta function  
on $r\in [0,+\infty)$ with the integration measure $\, \sim\! r^2\rd r$. 
Again, there are two families of eigenvalues due to the boundary conditions 
$f(r)=0$ and $f'(r)=0$ at $r=0$, which require the admissible values of 
$\delta$ to be roots of the Airy function $\mathtt{Ai}$ and its derivative. 
Hence, the spectrum of $\what{R}$ is discrete and bounded from below, 
and  the eigenvalues are labeled by the non-negative integers 
 $K\in\mathbb{Z}_+$
\begin{align} \label{est3d}
\begin{split}
&\mu_K =  \left(\frac{9 g^2\hbar^4\ka}{32} \right)^{1/3} |{\mathsf{ai}'_{K/2+1}}| 
\quad \; \; \mathrm{for\,even\,} K, 
\\
& 
\mu_K =  \left(\frac{9g^2\hbar^4\ka}{32} \right)^{1/3} |{\mathsf{ai}_{(K+1)/2}}| 
\quad \mathrm{for\,odd\,} K, 
 \end{split}
\end{align} 
where $\mathsf{ai}_M$ and  $\mathsf{ai}'_M$ denote the $M$-th root of the Airy function 
$\mathsf{Ai}$ and its derivative. 
Then the approximate estimation $\frac{1}{\ka}\what{H} \gtrsim \what{R}$ proves Theorem 1 independently from the result by B. Simon quoted above (though still using the insights from his paper \cite{simon83}). 

Moreover, from (\ref{est3d}) we obtain an estimation for the ground state 
of the DW Hamiltonian operator at $D=3$:  
\beq \label{322}
\left< \frac{1}{\ka}\what{H}  \right>_0 \gtrsim \mu_0 = \left(\frac{9g^2\hbar^4\ka}{32} \right)^{1/3} 
|{\mathsf{ai}'_1}| 
\eeq 
and the gap between  the first excited state and the ground state:
\beq \label{323}
\Delta\mu \approx \mu_1-\mu_0 = (|{\mathsf{ai}_1}|-|{\mathsf{ai}'_1}|)
\left(\frac{9g^2\hbar^4\ka}{32} \right)^{1/3} 
\approx 
         \mathtt{0.86}  \left({g^2\hbar^4\ka} \right)^{1/3} \,.
\eeq 
Note that, 
according to our conventions in (\ref{dwhop}),  
the gauge coupling constant $g$ in $3+1$ dimensions (i.e. the bare constant present in the Lagrangian) is given in the units of $1/\sqrt{\hbar}$, 
hence it is dimensionless in the units with $\hbar=1$. Correspondingly, the dimensionality of the r.h.s of (\ref{323}) is that of a mass. 
  Eq. (\ref{323}) tells us  that the  scale of the ultra-violet parameter $\ka$ introduced in  precanonical quantization is connected with the scale of the first massive excitation of quantum pure YM theory. Interestingly, the ground state of $\what{H}$ in (\ref{dwhop}) 
is independent of the ordering of the multiplicative operators $A$ and the differential 
operators $\der_A$ because of  the specific structure of the interaction term 
in the DW Hamiltonian operator and the antisymmetry of the structure constants. 

Let us also note that a comparison of the coefficient $\frac{Dg\hbar}{4}$ in front of the 
 $|\bB|$ term in (\ref{est4}) with the coefficient $\frac{g\hbar}{2}\widetilde{m}$ in front of the linear term in (\ref{est02a}) indicates that at $D>2$ it is not excluded that 
 there can be 
 colored mass excitations with nonvanishing quantum numbers $\widetilde{m}_{ij}$ which are lying below the estimated mass gap (\ref{323}) for the colorless excitations  
 with $\widetilde{m}_{ij}=0$.

\section{Conclusion}

The approach of precanonical quantization leads to the description 
of quantum pure Yang-Mills theory in terms of the Clifford-algebra-valued 
precanonical wave function on the space of Yang-Mills field variables 
$A^a_\mu$ and space-time coordinates $x^\mu$. Hence,  the  quantum YM field 
is understood as a section in the Clifford bundle over the bundle of gauge field 
components over space-time. This contrasts with the usual description of 
quantum fields in terms of the Schr\"odinger wave functionals or operator-valued 
distributions, or operator algebras. The precanonical wave function satisfies 
the covariant analogue of the Schr\"odinger equation defined on the aforementioned 
bundle, eq. (\ref{schreq}), which is a Dirac-like PDE on this bundle, with 
the mass term replaced by the DW Hamiltonian operator. 
 Note that,  
 unlike the quantization based on the canonical Hamiltonian formalism,  
  this approach treats all space-time dimensions on an equal footing. 
 Moreover, the construction of precanonical quantum field theory of YM fields is intrinsically 
 nonperturbative. 
We have demonstrated elsewhere \cite{my-ehrenfest} that this formulation is also 
consistent with a generalized version of the Ehrenfest theorem, i.e. the classical 
field equations are reproduced as the equations satisfied by the expectation values of precanonical 
operators \cite{my-ehrenfest}. 
Moreover, the standard quantum YM theory in the functional Schr\"odinger representation 
can be derived from our precanonical formulation in the 
limiting case of an infinite value of the ultraviolet parameter $\ka$ \cite{my-ym2016} 
(see also \cite{my-schrod}).  Thus the standard QFT, which requires a UV regularization, 
appears as a limiting case of the precanonical formulation which has the UV scale $\ka$ 
built in {\em ab initio}. 

The appearance of the DW Hamiltonian operator 
 in place of the mass term in the Dirac-like precanonical Schr\"odinger equation (\ref{schreq})
 indicates that its spectrum has to do with the spectrum 
of propagating excitations of the field, 
which one usually calls particles. This has motivated our interest in the spectrum of the DW Hamiltonian operator of pure YM theory. The discreteness of the spectrum, which we have proven for the collorless states, indicates that the propagating excitations 
in quantum YM theory are massive. Our consideration has lead to the estimation of the ground state of the DW Hamiltonian operator 
 and 
 the gap between the ground state and the first colorless excited state. Both expressions 
are\ $\sim\!g^{2/3}\ka^{1/3}$  in three spatial dimensions. 
This relates the scale of $\ka$,  up to a coefficient given by the bare gauge coupling $g$ in the Lagrangian, to the scale of the mass gap of the quantum nonabelian gauge theory under consideration. 
 
 Recently, a rough estimation of the scale of $\ka$ was obtained by us 
by a completely independent consideration based on  
the precanonical quantization of gravity \cite{my-qgr}. There,  $\ka$ appears 
in the ordering-dependent dimensionless 
combination with $\hbar$, 
Newton's $G$  and the cosmological constant $\Lambda$. A preliminary consideration has 
shown that the value of $\ka$ consistent with the observable values of the constants of nature 
is roughly at the  subnuclear scale. An order of magnitude coincidence of this estimation 
with the above estimation of the mass gap in $SU(2)$ quantum gauge theory  indicates, albeit preliminarily,  that $\ka$ is a fundamental scale rather than a kind of renormalization 
scale to be removed from the final results. 
Note that in spite of this indication to the 
fundamental scale at such a low energy, its existence in our formalism  does not contradict 
the current experimental evidence 
that the relativistic space-time holds at least till the $TeV$ energies, 
because the scale $\ka$ has been introduced in precanonical 
quantization without a manual modification  of the relativistic space-time at small distances. 

 \medskip 
 
\section*{Acknowlegdements} I thank Marek Czachor (Gda\'nsk) 
for his critical comments which have stimulated me to 
improve the arguments in an earlier version of the paper. 
I also thank 
Eckhard Hitzer (Tokyo) for a useful remark regarding the Clifford Analysis. 
This paper would not see the light of day 
without the hospitality of the School of Physics and Astronomy 
of the University of St Andrews and 24/7 availability of its facilities for research, which I gratefully appreciate. 



\end{document}